\documentclass[aps,prl,twocolumn,showpacs]{revtex4}
\usepackage{epsfig}
\usepackage{latexsym}
\usepackage{float}
\newfloat{widefig}{thp}{lop}

\begin{document}


\title{Effect of macromolecular crowding on the rate of
diffusion-limited enzymatic reaction}

\author{Manish Agrawal$^1$, S. B. Santra$^2$, Rajat Anand$^1$ and
Rajaram Swaminathan$^1$}

\affiliation{$^1$Department of Biotechnology, $^2$Department of Physics,\\
Indian Institute of Technology Guwahati, Guwahati-781039, Assam,
India.}

 
\begin{abstract} 
The cytoplasm of a living cell is crowded with several
macromolecules of different shapes and sizes. Molecular diffusion in
such a medium becomes anomalous due to the presence of macromolecules
and diffusivity is expected to decrease with increase in
macromolecular crowding. Moreover, many cellular processes are
dependent on molecular diffusion in the cell cytosol. The enzymatic
reaction rate has been shown to be affected by the presence of such
macromolecules. A simple numerical model is proposed here based on
percolation and diffusion in disordered systems to study the effect of
macromolecular crowding on the enzymatic reaction rates. The model
explains qualitatively some of the experimental observations. 
\end{abstract}

\maketitle

\section{Introduction}
The aqueous phase of cell cytoplasm is crowded with macromolecules
such as soluble proteins, nucleic acids and membranes
\cite{Fulton}. The influence of such crowding on biochemical reactions
inside physiological media are manifold \cite{MintonAP}. Due to
crowding, the average free energy $\mu$ of a nonspecific interaction
between any molecule in the medium and a crowding molecule may change
considerably which may influence the reaction activity $\gamma
=\exp(\mu/k_BT)$, where $k_B$ is the Boltzmann constant and $T$ is the
absolute temperature. Steric repulsion is the most fundamental of all
interactions between macromolecules in solution at finite
concentration and as an effect of such repulsion the macromolecules
occupy a substantial volume fraction in the cell interior
\cite{Zimmermann}. Significant volume fraction of macromolecules in
the medium imposes a constraint on introducing any new
macromolecule. As a consequence of crowding, macromolecular
association reactions become increasingly favorable. Because of
crowding, the molecular diffusion in the medium is expected to be
anomalous \cite{Saxton}. The effect of macromolecular crowding on
different kinetic steps of enzyme catalysis such as formation of
enzyme-substrate complex and enzyme-product complex were analyzed
through different equilibrium thermodynamic models\cite{Minton}. A
number of approaches have been proposed to study the effects of
macromolecular crowding on the reaction kinetic rate laws such as the
law of mass action \cite{Minton}, fractal like kinetics
\cite{fractalk}, the power law approximation \cite{power}, stochastic
simulation\cite{stochastic} and lattice gas
simulation\cite{latticeg}. In these analytic and numerical models, the
influence of macromolecular crowding on both equilibrium
thermodynamics and reaction rates were addressed and it was observed
that the rate decays exponentially with time as expected in
equilibrium kinetics. The influence of macromolecular crowding on the
enzymatic reaction rates has been investigated experimentally using a
variety of crowding agents \cite{expt1}. These studies have also
indicated a significant influence of crowding on the rate parameters
of the enzymatic reaction. It was found that the effect of crowding on
the enzymatic reaction may be different depending on whether the
product formation in the enzyme reaction is limited by the diffusional
encounter of substrate and enzyme or the formation of the transition
state complex, an association of enzyme and substrate. Moreover,
molecular diffusion is known to be the major determinant of many
cellular processes and plays a key role in cell metabolism where the
encounter of the free substrate with an active site of the enzyme is
often the rate determining step. However, how the kinetics of an
enzymatic reaction is dependent on the size and concentration of the
crowding macromolecules is still not fully understood. The
macromolecular crowding till date remains under appreciated and
neglected aspect of the intracellular environment \cite{ellis}. It is
hence essential to understand the experimental observations from
microscopic origin.

In this paper, an approach based on non-equilibrium dynamics of
enzymatic reactions in the diffusion limited regime is considered. The
aim is to understand qualitatively the influence of inert
macromolecular crowding on the diffusion limited enzymatic reactions
governed by non-equilibrium thermodynamics. A simple numerical model
in two dimensions ($2d$) based on molecular diffusion in disordered
systems coupled with enzymatic reaction is proposed here. The
disordered system is modeled by percolation clusters \cite{perco}. It
is predicted that the rate of a diffusion-limited enzyme-catalyzed
reaction will experience a monotonic decrease with increase in the
fractional volume occupancy of the crowding agent. The model explains
qualitatively certain experimental observations.

\section{The Model}
In brief, the enzyme kinetic reaction in the cell cytoplasm can be
described as substrate molecules diffusing through crowding
macromolecules and binding to the active site of the freely floating
enzymes. Subsequently a product is formed if the reaction is
energetically favorable and this product diffuses through the same
crowd of macromolecules. The classical Michaelis-Menten equilibrium
enzyme kinetic reaction is given as \cite{ekinetics}
\begin{equation}
\label{eq1}
E+S\rightleftharpoons ES \rightarrow E+P
\end{equation} 
where $E$ represents enzyme, $S$ represents substrate, $P$ represents
product and $ES$ is the intermediate enzyme-substrate complex.

In the present model, the reaction is limited by diffusion only and
the formation of the transition state complex $ES$ is not taken into
account. The conversion of substrate to product is also assumed to be
instantaneous. Note that diffusion has the slowest time scale in this
problem. Hence, the above enzymatic reaction reduces to an
irreversible one as
\begin{equation}
\label{eq2}
E+S\rightarrow E+P.
\end{equation} 
The final equilibrium state corresponds to conversion of all
substrates to products. A Monte Carlo (MC) algorithm has been
developed to study diffusion limited enzymatic reaction as in
Eq.\ref{eq2} in the presence of inert macromolecules. The algorithm is
developed on the $2d$ square lattice of size $L\times L$. For
simplicity, the motion of the macromolecules is ignored and these act
as immobile and inert obstacles. The inert obstacles do not interact
with either among themselves or with the substrate or product. The
obstacles ($O$), enzyme ($E$), substrate ($S$) and product ($P$) are
all represented as point particles in this model. It is also assumed
that there exists only one immobile enzyme in the whole system. The
enzyme is placed at the center of the lattice. After placing the
enzyme, the obstacles and the substrates are distributed randomly over
the lattice sites with their specified concentrations $C_O$ and $C_S$
respectively. A random number $r$ is called from a uniform
distribution of random numbers between $0$ and $1$ corresponding to
each lattice site. If $r\le C_S$, the site is occupied with a
substrate and if $C_S< r\le a_f$ the site is occupied with an obstacle
where $a_f=C_S+C_O$ is the area fraction. The excluded volume
condition is maintained, {\em i.e.}, at any instant of time one
lattice site cannot be occupied by more than one molecule of the same
or different species. The substrate molecules diffuse through the
space not occupied by the obstacles which will be referred as empty
space later. As soon as a $S$ reaches $E$, a product $P$ is produced
with unit probability. The diffusion of substrate or product in the
system is modeled by simple random walk in presence of obstacles or
disorder. At each MC time step, all the random walkers (all $S$ and
$P$) make an attempt to move to one of their nearest neighbors. The
destination site, a site out of the four neighbors, of a random walker
is chosen calling a random number $r$ uniformly distributed between
$0$ and $1$. With respect to the present site, the destination site is
going to be on the left if $0 < r \le 1/4$, it is at the top if $1/4 <
r \le 1/2$, it is on the right if $1/2 < r \le 3/4$, and it is at the
bottom if $3/4 <r \le 1$. The destination site could be either empty
or occupied by $S$, $P$, $O$ or $E$. Depending on the status of the
destination site, there are then four possibilities: ($a$) if the
destination site is empty, the present $S$ or $P$ moves to the
destination site, ($b$) if the destination site is occupied by a $S$
or $P$, $S$ or $P$ remains on the same site, ($c$) if the destination
site is occupied by an $O$, $P$ or $S$ also remains on the same site,
and ($d$) if the destination site is occupied by the enzyme $E$, $P$
remains on the same site but $S$ is converted to $P$ with unit
probability. If all the molecules of $S$ and $P$ are checked for an
attempt of motion, time $t$ (the MC time step) is increased to $t+1$.
To ensure percolation of the substrate molecules, the maximum area
fraction $a_f=C_S+C_O$ is taken as $0.4$, far below the percolation
threshold. Note that, the percolation threshold on the square lattice
is $\approx 0.59$ \cite{perco}. Note that, the present non-equilibrium
diffusion limited enzymatic reaction model is substantially different
from that of lattice gas model incorporating equilibrium reaction
rates proposed by Schnell and Turner \cite{latticeg} which leads to an
unusual equilibrium constant equal to zero in the crowed environment
\cite{bajzer}.

\begin{figure*}
\centerline{\hfill \psfig{file=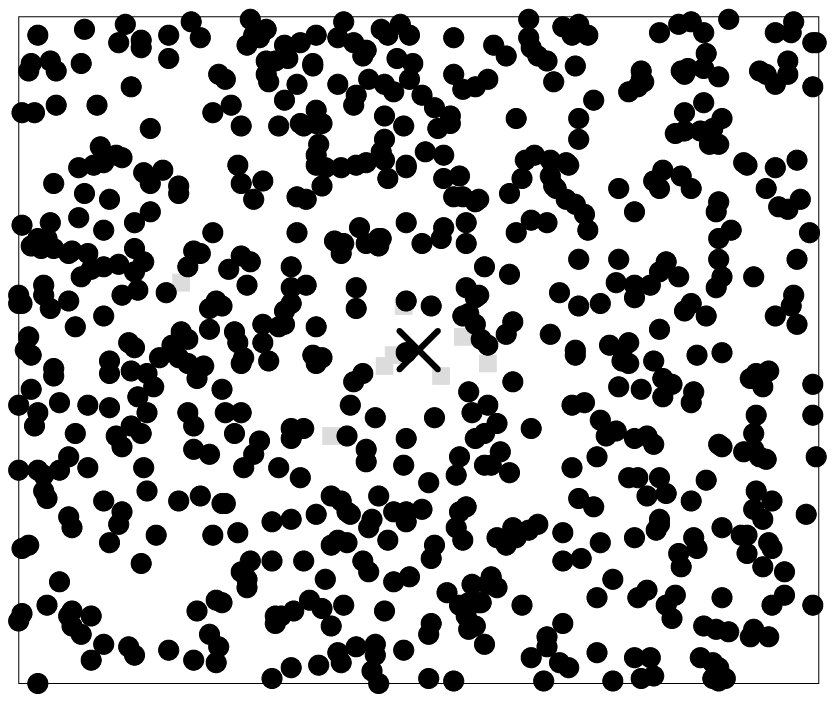,width=0.25\textwidth} \hfill
  \psfig{file=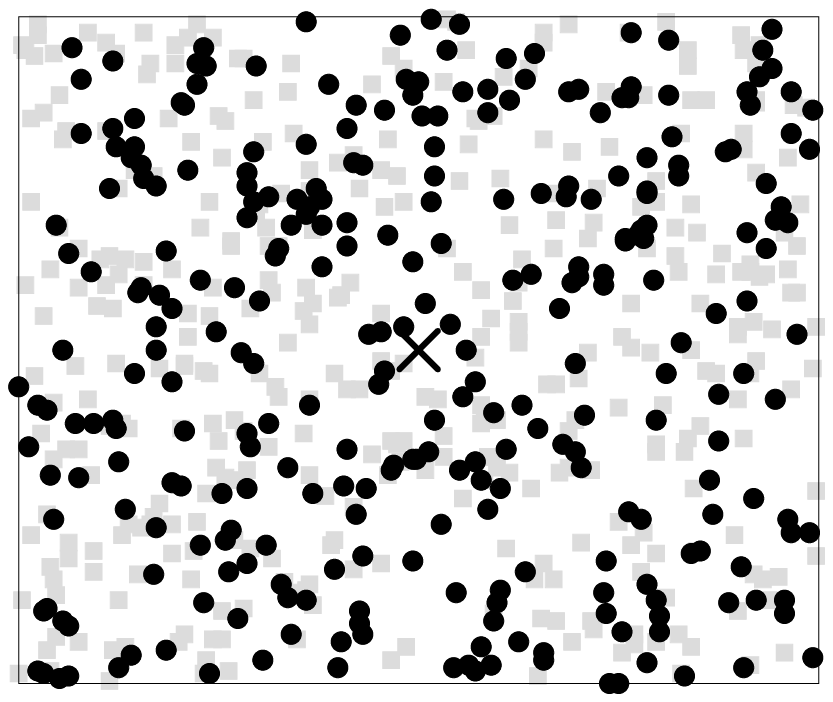,width=0.25\textwidth} \hfill
  \psfig{file=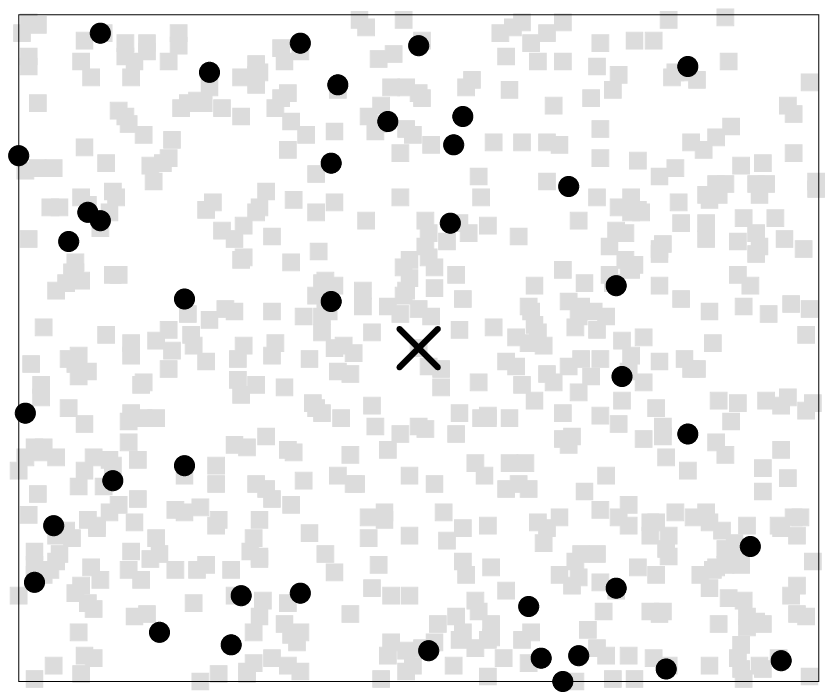,width=0.25\textwidth} \hfill}

\medskip
\centerline{\hfill ($a$) $t=2^{12}$ \hfill\hfill ($b$) $t=2^{18}$
  \hfill\hfill ($c$) $t=2^{20}$ \hfill }
\medskip
\caption{\label{fig:pic1} The system morphology on a $256\times 256$
  square lattice is shown at three different times $(a)$ $t=2^{12}$,
  $(b)$ $t=2^{18}$ and $(c)$ $t=2^{20}$ for substrate concentration
  $C_s=0.01$ and area fraction $a_f=C_S+C_O=0.1$ ($C_O=0.09$). The
  black dots represent the substrates and the gray boxes represent the
  products. For clarity obstacles are not shown. The enzyme is
  represented by a cross at the center of the lattice. Products are
  formed due to the enzymatic reaction and in the long time limit
  almost all the substrates are converted into products.}
\end{figure*}

Cyclic boundary condition has been applied in the motion of $S$ and
$P$. The simulation has been performed upto $10^6$ MC time steps on a
$256\times 256$ square lattice. The data are averaged over $100$
ensembles. The time evolution of the system morphology for $a_f=0.1$
with $C_S=0.01$ is shown in Fig.\ref{fig:pic1} at three different
time. The black dots represent the substrates and the gray boxes
represent the products. For clarity obstacles are not shown. It can be
seen that the initial black dots are converted to gray boxes at the
end. That means, the substrate molecules are diffusing, reacting with
the enzyme, and are getting converted into products. In time, almost
all the substrate molecules are converted to products and the product
molecules also diffuse and spread all over the space uniformly. Lin
and coworkers \cite{lin} simulated some elementary kinetic reactions
like $A+B\rightarrow 0$ with no obstacles under reflective boundary
condition and observed Zeldovich crossover (segregation of $A$ and
$B$) \cite{redner}.  Such segregation is not observed with periodic
boundary condition in the present simulation. Effect of impenetrable
boundary on diffusion limited reaction like $A+A\rightarrow 0$ leads
to different behavior depending on different boundary conditions
\cite{kafri}. 

\section{Results and discussion}
Classical diffusion of a tracer particle in disordered systems has
already been studied extensively and the results are well understood
\cite{r:diff}. Generally the diffusion is modeled by random walk and
the disordered system is modeled by spanning percolation clusters
\cite{perco}. For studying diffusion, a quantity of interest is the
root mean square (rms) distance $r(t)$ covered by the diffusing
particle in time $t$. The rms distance $r(t)$ in $2d$ is given by
\begin{equation}
\label{eq:diff}
r^2(t) = 4{\mathcal D}\times t^{2k}
\end{equation}
where $\mathcal D$ is the diffusivity of the system. The exponent $k$
has a value $1/2$ for diffusion on a regular lattice in the
$t\rightarrow \infty$ limit. On the percolation cluster, diffusion is
found to be anomalous and the value of $k$ becomes less than
$1/2$\cite{r:diff}. The enzyme kinetic reaction inside a cell
cytoplasm involves ($i$) diffusion of a large number of substrate
molecules through the random structure of inert macromolecules, ($ii$)
reaction with the enzyme to have products, and ($iii$) finally
diffusion of products from the enzyme through the same macromolecular
crowding. The diffusion process involved here is then a collective
motion of a large number of particles in presence of disorder which is
a complicated process than diffusion of a single tracer particle in a
disordered medium. Self-diffusion is expected to play a nontrivial
role along with the diffusion of $S$ or $P$ in presence of disorder in
these systems. In order to check whether the enzyme kinetic reaction
considered here is diffusion limited or not, one needs to analyze the
the diffusive behavior of either the substrates or the products. To
calculate the average diffusion length of the product particles, the
coordinates $\{x_i(t),y_i(t)\}$ of each product $i$ is recorded with
time $t$. Time is measured starting from the birth of a product. The
rms distance $r(t)$ traveled in time $t$ is then calculated as
\begin{equation}
\label{drms}
r^2(t) = \frac{1}{N_p(t)} \sum_{i=1}^{N_p(t)}\left[ \{x_0-x_i(t)\}^2 +
  \{y_0-y_i(t)\}^2 \right]
\end{equation}   
where $(x_0,y_0)$ is the coordinate of the enzyme at the center of the
lattice, $N_p(t)$ is number of products of age $t$. The data is then
sample averaged over $100$ ensembles.

\begin{figure}[!h]
\centerline{\hfill \psfig{file=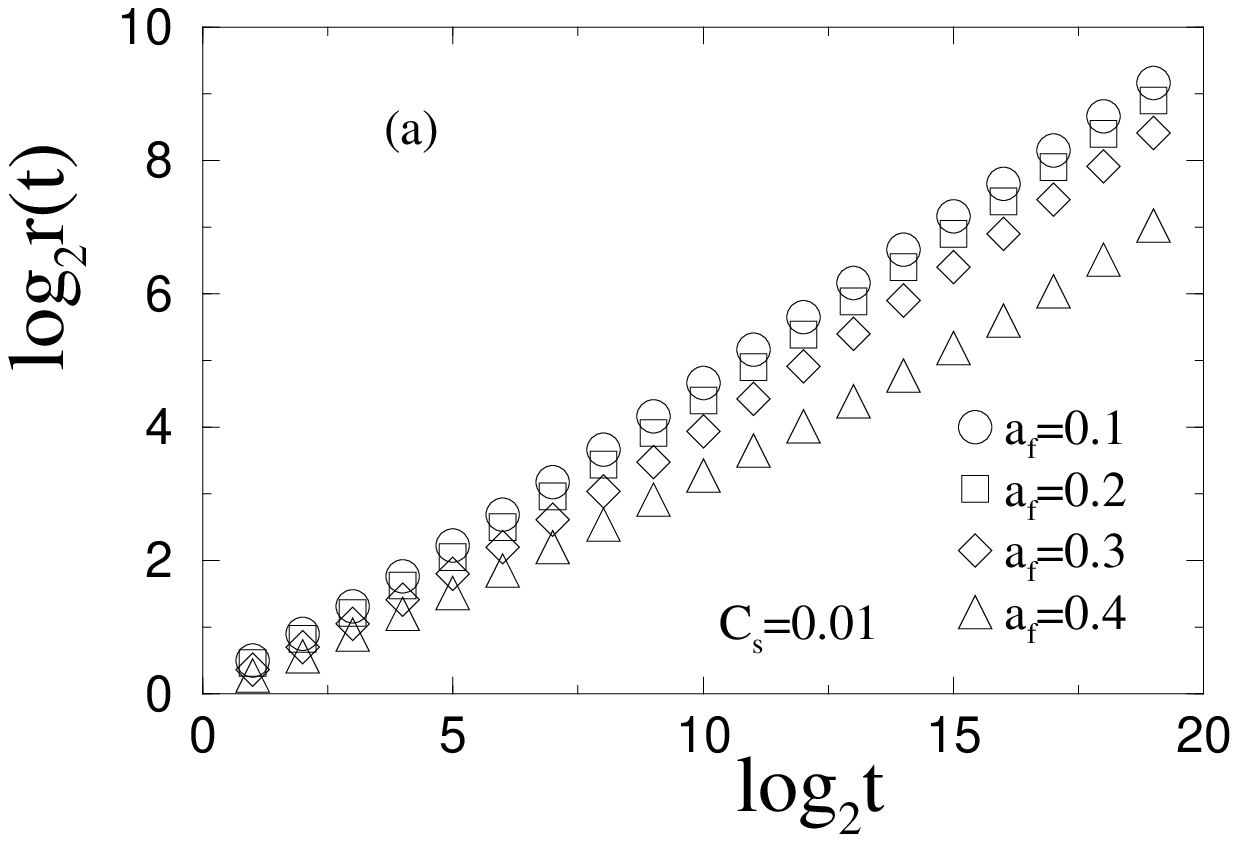,width=0.4\textwidth} \hfill}
\centerline{\hfill  \psfig{file=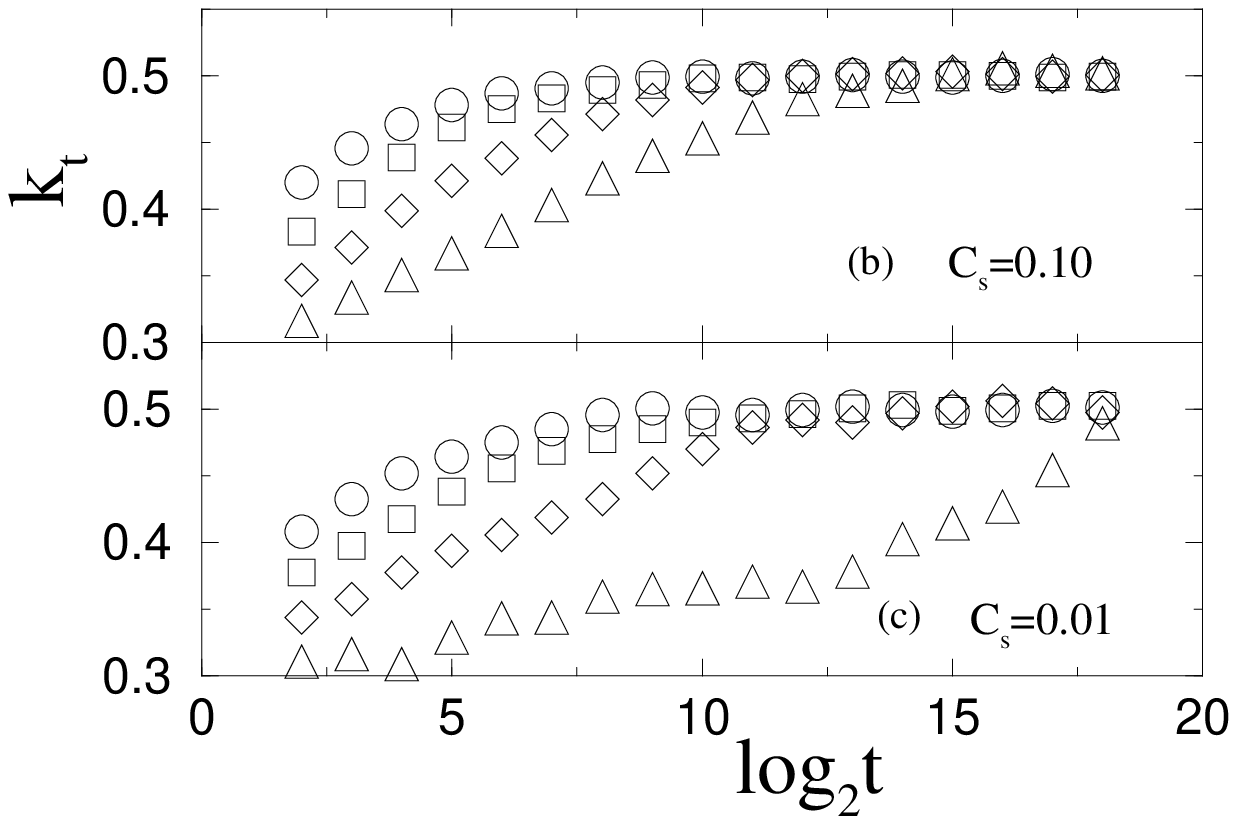,width=0.4\textwidth } \hfill}
\medskip
\caption{($a$) Plot of diffusion length $r(t)$ against time $t$ in
  double logarithmic scale for different are fractions $a_f$ keeping
  substrate concentration at $C_S=0.01$. $(b)$ and $(c)$ Plot of the
  local exponent $k_t$ versus time $t$ for $C_S=0.1$ and $C_S=0.01$
  respectively. The same symbol set of $(a)$ has been used in $(b)$
  and $(c)$ for different area fractions $a_f=C_S+C_O$. }
\label{fig:diff}
\end{figure}

In Fig.\ref{fig:diff}($a$), $r(t)$ is plotted against time $t$ in
double logarithmic scale for different area fractions $a_f=C_S+C_O$
keeping the substrate concentration constant at $C_S=0.01$. It can be
seen that the magnitude of the diffusion length decreases with
increasing $a_f$. The collective motion of the particles is then
affected more and more by the presence of increasing inert
macromolecules in the system. However, in order to check the diffusive
behavior of the particles one needs to estimate the exponent $k$
defined in Eq.\ref{eq:diff}. The local slope
$k_t=d\log_2r(t)/d\log_2t$ of the curve $\log_2r(t)$ versus $\log_2t$
is determined by employing central difference method. In
Fig.\ref{fig:diff}($b$) and $(c)$, $k_t$ is plotted against time $t$
for two different substrate concentrations $C_S=0.10$ $(b)$ and
$C_S=0.01$ $(c)$ for the same set of area fractions $a_f$ as in
Fig.\ref{fig:diff}($a$). The value of $k_t$ saturates to $1/2$
starting from a smaller value as $t$ tends to a large value. Thus, a
crossover from sub-diffusive to diffusive behavior has occurred for
all area fractions in the long time limit. In the case of low
substrate concentration $C_s=0.01$ and high area fraction $a_f=0.4$,
$k_t$ shows certain anomalous behavior. Note that, at this parameter
regime the macromolecular concentration is $0.39$ which is just below
$1-p_c\approx 0.41$ since the percolation threshold for a $2d$ square
lattice is $p_c \approx 0.59$. The empty sites provides the
connectivity for the substrate molecules all over the
lattice. However, $p_c$ is defined on a infinitely large system. For a
smaller system, even at the concentration of $0.39$ the connectivity
of empty sites may be lost for some of the ensembles
considered. Consequently, the product may be trapped in a localized
region around the enzyme and this may be the reason behind the
anomalous behavior observed in this parameter regime.

\begin{figure}[!h]
\centerline{\hfill \psfig{file=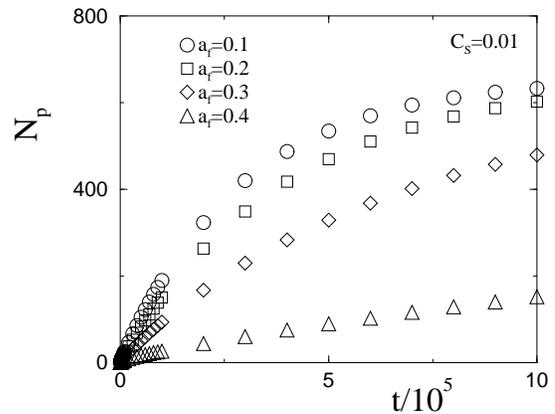,width=0.4\textwidth} \hfill}
\medskip
\caption{ Plot of number of products $N_P$ versus time $t$ for
  different area fractions $a_f=C_S+C_O$ keeping substrate
  concentration constant at $C_S=0.01$.  }
\label{fig:np}
\end{figure}

Since the parameter regime here is limited by diffusion, the enzyme
kinetic reaction is then expected to be diffusion limited. Due to the
enzyme kinetic reaction (given in Eq.\ref{eq2}) the substrates are
converted to products in time with unit probability on their
encounter. In order to characterize the enzyme kinetic reaction, the
number of products $N_P$ are counted as function of time $t$, the MC
time step, for different substrate concentrations $C_S$ and area
fractions $a_f=C_S+C_O$. In Fig.\ref{fig:np}, the product numbers
$N_P$ is plotted against time $t$ for different area fractions $a_f$
with $C_S=0.01$. Initially, $N_P$ increases linearly, then slows down
and finally saturates in the long time limit. For low area fraction,
it can be seen that the reaction is almost complete {\em i.e.}; most
of the substrates given initially, $N_S(0)=C_S\times L^2\approx 655$,
are converted to products exponentially as in classical equilibrium
Michaelis-Menten kinetics though in the present model a
non-equilibrium kinetics is considered. However, note that there is a
considerable decrease in the product yield with increase in area
fraction and their profiles are found not to follow an exponential
increase. It has already been predicted by numerical simulations that
classical Michaelis-Menten kinetics may not apply to enzymatic
reactions in crowded media\cite{berry}. In a $1d$ model of reaction
diffusion with disorder, Doussal and Monthus \cite{doussal} also found
large time decay in the species density via real space renormalization
group calculations. The macromolecular crowding then could have a
considerable and nontrivial effect on the enzymatic reaction rate.

\begin{figure}[!h]
\centerline{\hfill \psfig{file=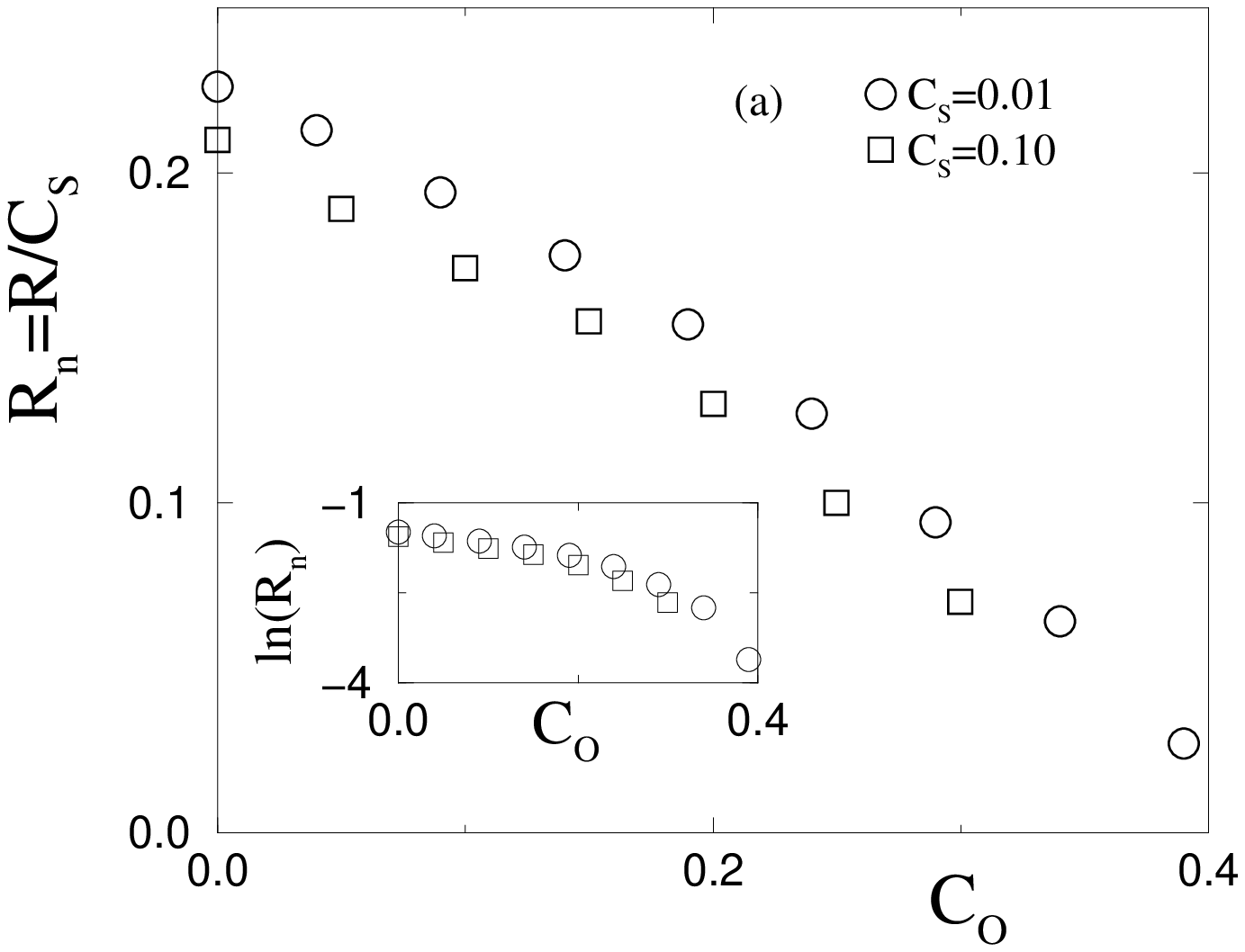,width=0.35\textwidth} \hfill}
\centerline{\hfill \psfig{file=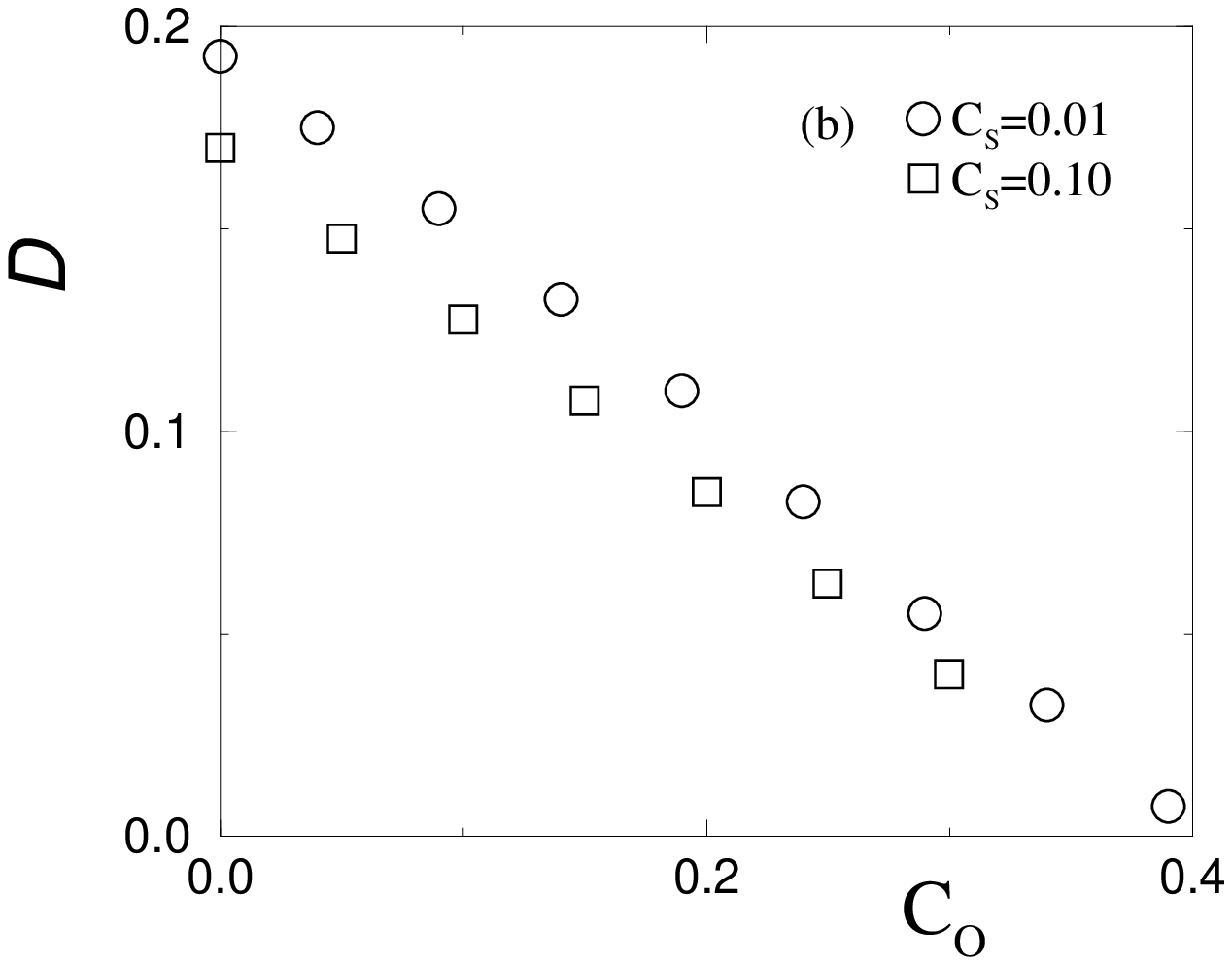,width=0.35\textwidth} \hfill}
\medskip
\caption{($a$) Plot of normalized reaction rate $R_n=R/C_S$ against
   $C_O$ for two different $C_S$ values $0.01$ (circles) and $0.1$
   (squares). $\ln(R_n)$ is plotted against $C_O$ for the same $C_S$
   values in the inset. The same symbol set for different $C_S$ values
   is used. $(b)$ Plot of diffusivity $\mathcal D$ against $C_O$ for
   $C_S=0.01$ and $C_S=0.1$. The same symbol set of ($a$) is used.}
\label{fig:rr}
\end{figure}

Initial rate of enzymatic reactions determines most of the molecular
process and thus is an important quantity to estimate. Since
non-equilibrium enzymatic reaction is considered here, the reaction
rate $R$ is defined as the ratio of the number of products $N_P$ to
time $t$ for $10\%$ conversion of the substrates. $R$ is then sample
averaged. A similar analysis has also been performed for $N_P$ versus
$t$ plots corresponding to $C_S=0.1$ for different area fractions
$a_f$. In Fig.\ref{fig:rr}($a$), the normalized reaction rate
$R_n=R/C_S$ is plotted against obstacle concentration $C_O$ for two
different substrate concentrations $C_S=0.01$ (circles) and $C_S=0.1$
(squares). Note that, area fraction $a_f=C_S+C_O$ is not a good
parameter to study the reaction rate since $a_f$ will remain finite
for finite $C_S$ even at $C_O=0$. In the inset, $R_n$ is also plotted
against $C_O$ in semi logarithmic scale. There are few things to
notice. First, the reaction rate is decreasing with the increase in
obstacle concentration $C_O$ in a nonlinear fashion. Second, the
reaction rates are different for a particular $C_O$ even after
normalizing by the substrate concentration $C_S$. Third, there is a
monotonic decrease of $\ln(R_n)$ for small $C_O$ and deviates from
linear decrease for large $C_O$. The decrease in reaction rate with
increasing crowding concentration is expected and also observed in
experiments \cite{expt1,rs}. However, the dependence of the rate on
the crowding concentration is different form the prediction made by
Minton \cite{Minton} in the transition state as well as diffusion
limited enzymatic reaction in which a hump in the $\ln(R_n)$ versus
$C_O$ curve is expected for an intermediate $C_O$. Fourth, the
normalized reaction rate is going to zero as $C_O$ approaches $1-p_c
\approx 0.41$. Beyond $C_O=0.41$, the obstacles could block the
spanning clusters of the empty sites. Consequently the enzymatic
reaction will be localized and the reaction rate is expected to go to
zero.

The above observations can qualitatively be understood in terms of
diffusion and percolation phenomena. As $C_O$ increases, diffusivity
is expected to decrease because of the crowding due to obstacles. The
influence of macromolecular crowding on the diffusion of solutes has
been investigated in recent experiments utilizing different crowding
agents and a reduced solute diffusion coefficient was observed with
increasing size and concentration of crowding macromolecules
\cite{expt2}.  An estimate of diffusivity ${\mathcal D}
=(dr^2(t)/dt)/4$ (as given in Eq.\ref{eq:diff}) has been made
utilizing the data of diffusion length $r(t)$ for different sets of
substrate $(C_S)$ and obstacle $(C_O)$ concentrations. In
Fig.\ref{fig:rr}($b$), $\mathcal D$ is plotted against $C_O$ for
$C_S=0.01$ (circles) and $C_S=0.10$ (squares). Like reaction rate,
diffusivity $\mathcal D$ is also studied as a function of obstacle
concentration $C_O$ instead of $a_f$. It can be seen that $\mathcal D$
also decreases with $C_O$ in a nonlinear fashion. First of all, it is
interesting to note that the whole dependence of $R_n$ on $C_O$ is is
in accordance with the behavior of $\mathcal D$ with $C_O$. The
enzymatic reaction rate in this parameter regime is therefore mostly
governed by diffusion and can be considered a purely diffusion limited
enzymatic reaction. It is important now to consider the low $C_O$
values, especially the case of $C_O=0$. For low $C_O$ values,
$\mathcal D$ is slightly less for $C_S=0.1$ than that of $C_S=0.01$
for the same $C_O$. This slight decrease in $\mathcal D$ is due to
diffusion through the self crowding at higher $C_S$. On the other
hand, the reaction rate at zero obstacle concentration is expected to
be proportional to $C_S$ and $\mathcal D$ and $R$ can be obtained as
$R\approx C_S\times{\mathcal D}$. It can be seen that the normalized
reaction rate $R_n$ obtained here is very close to the corresponding
values of $\mathcal D$ at $C_O=0$ for both the $C_S$ values. At
$C_O=0$, the self diffusion of the substrate molecules eventually
determines the reaction rates and might be responsible for a slight
decrease in $R_n$ for $C_S=0.1$ with respect to $C_S=0.01$ as seen in
Fig.\ref{fig:rr}($a$). The effect of $C_S$ in absence of obstacles has
been verified numerically for several higher values of $C_S$ and a
considerable effect of self-crowding has been observed on the reaction
rate as well as on diffusivity. Note that, $R_n$ values are slightly
greater than $\mathcal D$ for almost all values of $C_O$ as it can be
seen by comparing Fig.\ref{fig:rr}($a$) and $(b)$. This might have
happened firstly due to the fact that the initial yield occur only
from the locally available substrate molecules. The diffusion length
of these substrate molecules are very less in comparison to the
expected diffusion length. Secondly, one should note that the initial
reaction rate for a given $C_O$ has to be calculated keeping the
substrate concentration $C_S$ fixed. However, in the present model the
substrate concentration is decreasing with time as the substrates are
being converted into products. The effect will be predominant for low
$C_S$ and small system size. Consequently the rate determination will
be erroneous in the $t\rightarrow 0$ limit due to low yield. Hence,
extreme care has to be taken in determining the initial reaction rate.
The enzymatic reaction considered here is the completely diffusion
limited and the results obtained are explainable in terms of diffusion
in disordered systems. It is therefore intriguing to note that such a
simple model of enzymatic reaction based on diffusion and percolation
phenomena only, is able to explain qualitatively the experimental
observations \cite{expt1,rs} as well as results obtained in
complicated models
\cite{Minton,fractalk,power,stochastic,latticeg}. Hence, diffusion is
observed to be playing the crucial role in determining the enzymatic
reaction rates.

It should be emphasized here that enzymatic reactions occur in
$3$-dimensional space in living systems whereas the simulation is
performed in $2$-dimensions here. The simulation results obtained here
agree qualitatively with the experimental observations and it is
expected that the features of the model will be retained in higher
dimensions also. The main difficulty in $3d$ simulation is in parallel
updating of a large number of substrate and product molecules during
time evolution through a large number of MC time steps. Time required
for the full conversion of substrate to product increases
exponentially with the number of molecules ($N_S=C_S\times L^d$) which
strongly depends on the dimensionality of space for a fixed substrate
concentration. However, for quantitative comparison of the results
obtained in simulation with that of experiments, the model must be
extended to three dimensions.

The biochemical events in the densely crowded mitochondrial matrix,
the site for TCA cycle and fatty acid oxidation pathway are largely
governed by large macromolecules of various sizes. It is thus
important to investigate the influence of crowding as exerted by
macromolecules of different sizes. A decrease in reaction rate has
been observed in experiments for varying obstacle sizes keeping the
obstacle concentration constant\cite{rs}. It seems that the complex
interaction between obstacles and the substrate is size dependent and
might be governing the enzymatic reaction rate. It is expected that
the diffusion of substrates across large macromolecules might be slow
due to the complex interactions with the obstacles. In the present
model of enzymatic reaction, this complex interaction between obstacle
and substrate may be incorporated by introducing a residence time for
the substrate molecules at each encounter with the obstacle. A slowing
down in the reaction rate with increasing residence time has been
observed in the simulation in accordance with the experimental results
\cite{rs}. The details will be reported elsewhere.

\section{Summary}

The effect of macromolecular crowding on the enzymatic reaction rates
has been modeled by a MC algorithm based on diffusion and percolation
phenomena. The substrates, products, obstacles and enzyme all are
represented by point particles. A single immobile enzyme is considered
and placed at the center of the lattice. The obstacles and the
substrates are distributed randomly with their specific concentrations
following a uniform distribution of random numbers between $0$ and
$1$. The obstacles remain immobile throughout the simulation. It is
found that the reaction is solely diffusion limited under these
conditions. The diffusion of substrates and products are modeled by a
collective random walk. The products form gradually and subsequently
almost all the substrates are converted into products after a long
time. The initial reaction rate has been estimated for different
substrate and obstacle concentrations. The normalized reaction rate
has a nonlinear dependence on the obstacle concentration and found
slightly dependent on the substrate concentration. The dependence of
reaction rate on the substrate as well as obstacle concentrations is
then qualitatively understood with the help of diffusion and
percolation theory. The results qualitatively explains the
experimental observations.


\begin{thebibliography}{10}
\bibitem{Fulton} A B Fulton,  Cell {\bf 30}, 345 (1982).

\bibitem{MintonAP} S. B. Zimmerman and A. P.  Minton,
  Annu. Rev. Biophys. Biomol. Struct. {\bf 22}, 27 (1993);
  A. P. Minton, J. Biol. Chem. {\bf 276}, 10577 (2001).

\bibitem{Zimmermann} S. B. Zimmermann and S. O. Trach,
J. Mol. Biol. {\bf 222}, 599 (1991).

\bibitem{Saxton} M. J. Saxton,  Biophys. J. {\bf 66}, 394 (1994);
  {\bf 70}, 1250 (1996).

\bibitem{Minton} A. P. Minton, Biopolymers {\bf 20}, 2093 (1981).


\bibitem{fractalk} R. Kopelman, J. Stat. Phys. {\bf 42}, 185 (1986);
  Science {\bf 241}, 1620 (1988). 

\bibitem{power} M. A. Savageau, J. Theor. Biol. {\bf 176}, 115 (1995).

\bibitem{stochastic} D. T. Gillespie, Physica A {\bf 188}, 404 (1992);
  T. B. Kepler and T. C. Elston, Biophys. J. {\bf 81}, 3116 (2001).

\bibitem{latticeg} S. Schnell and T. E. Turner,
  Prog. Biophys. Mol. Biol. {\bf 85}, 235 (2004).

\bibitem{expt1} T. C. Laurent, Eur. J. Biochem. {\bf 21}, 498 (1971);
  A. P. Minton and J. Wilf, Biochemistry {\bf 20}, 4821 (1981);
  J. R. Wenner and V. A. Bloomfield, Biophys. J. {\bf 77}, 3234
  (1999); N. Assad, J. B. F. N. Engberts, J. Am. Chem. Soc. {\bf
  125}, 6874 (2003).

\bibitem{ellis} R. J. Ellis, Trends Biochem. Sci. {\bf 26}, 597
  (2001). 

\bibitem{perco} D. Stauffer and A. Aharony, {\em Introduction
  to percolation theory}  (Taylor and Francis, London) (1994).

\bibitem{ekinetics} K.B. Taylor, {\em Enzyme Kinetics and Mechanisms},
(Kluwer Academic Publishers, The Netherlands) (2002).

\bibitem{bajzer} Z. Bajzer, M. Huzak, K. Neff and F. G. Prendergast,
  Croat. Chem. Acta {\bf 79}, 437 (2006).

\bibitem{lin} A. Lin, R. Kopelman and P. Argyrakis, Phys. Rev. E {\bf
  53}, 1502 (1996).

\bibitem{redner} F. Leyvraz and S. Redner, Phys. Rev. A {\bf 46}, 3132
  (1992).

\bibitem{kafri} Y. Kafri and  M. J. E. Richardson, J. Phys. A {\bf
  32}, 3253 (1999). 

\bibitem{r:diff} P. G. de Gennes, La Recherche, {\bf 7}, 916 (1976);
  S. Havlin and D. Ben-Avraham, Adv. Phys. {\bf 36}, 695 (1987);
  S. B. Santra and W. A. Seitz, Int. J. Mod. Phys. C {\bf 11}, 1357
  (2000).

\bibitem{berry} H. Berry, Biophys. J. {\bf 83}, 1891 (2002).

\bibitem{doussal} P. Le Doussal and C. Monthus, Phys. Rev. E {\bf 60},
1212 (1999).

\bibitem{expt2} R. Swaminathan, C. P. Hoang and A. S. Verkman,
  Biophys. J. {\bf 72}, 1900 (1997); A. S. Verkman, Trends
  Biochem. Sci. {\bf 27}, 27 (2002); E. Dauty and A. S. Verkman,
  J. Mol. Recognit. {\bf 17}, 441 (2004).

\bibitem{rs} L. Homchaudhuri, N. Sarma and R. Swaminathan, Biopolymers
  {\bf 83}, 477 (2006).
\end{thebibliography}
\end{document}